\documentclass[onecolumn, superscriptaddress, showpacs]{revtex4-1} 
\usepackage{graphicx}
\usepackage{epstopdf}
\usepackage{amsmath}
\usepackage{color}

\begin{document}
\title{Kinetic roughening of a soft dewetting line under quenched disorder - a numerical study}

\author{B. Tyukodi}
\affiliation{Babe\c{s} - Bolyai University, Department of Physics \\ Cluj-Napoca, Romania}
\affiliation{Edutus College,\\ Tatab\'anya, Hungary}
\affiliation{Laboratoire PMMH, UMR7636 CNRS/ESPCI/Universit\'{e} Paris 6 UPMC \\Paris, France}
\author{Y. Brechet}
\affiliation{French Alternative Energies and Atomic Energy Commission, \\ Paris, France}
\affiliation{Grenoble Institute of Tehnology, SIMAP, \\ St. Martin d'Heres, France}
\author{ Z. N\'eda}
\affiliation{Babe\c{s} - Bolyai University, Department of Physics \\  Cluj-Napoca, Romania}
\affiliation{Edutus College,\\ Tatab\'anya, Hungary}


\begin{abstract}
A molecular-dynamics type simulation method, which is suitable for investigating the dewetting dynamics of thin and viscous liquid layers, is discussed. The efficiency of the method is exemplified by studying a two-parameter depinning-like model defined on inhomogeneous solid surfaces. The morphology and the statistical properties of the contact line is mapped in the relevant parameter space, and as a result critical behavior in the vicinity of the depinning transition is revealed. The model allows for the tearing of the layer, which leads to a new propagation regime resulting in non-trivial collective behavior. The large deformations observed for the interface is a result of the interplay between the substrate inhomogeneities and the capillary forces. 
\end{abstract}

\maketitle

\section{Introduction}
Contraction of thin liquid layers on solid surfaces due to dewetting or
drying is a common phenomenon. It is observable for instance, on plants' leafs as the
water breaks up into small droplets, in non-sticking pans as the oil layer
shrinks or on an outdoor oil-polluted surface after rain. Another well-know example is the
contraction of the liquid layer covering the eyeball, the characteristic time
scale of a complete contraction being the time elapsed between two successive
blinks \cite{cornea1, cornea}. Dewetting plays an important role in the tire
industry as well: when the contraction of the wetting layer on the tire's groove is too
slow, aquaplaning is more likely to occur \cite{PRL_Martin,
PRB_Persson, EPJE_Persson}. Dewetting is also important in the lubricant
manufacturing, however in this case exactly the opposite effect is desired: the
more a lubrifient remains on the surface of sliding pieces, i. e. the larger its
contraction time, the better.

Along with the development of the polymer industry, contraction of polymer
films started to gain interest \cite{PRL_Geiger,  vilmin, Xue}.
Dewetting  turned out to be a useful investigative tool for determining
various rheological and interfacial properties of thin polymer films due to the fact that 
molecular properties are reflected in the macroscopic shape of the
solid-liquid-gas triple interface \cite{EPJ_Reiter}.

In other cases, liquids are used as carriers for certain substances
(nanoparticles, for example), thus dewetting eventually accompanied by drying on
rough surfaces of such solutions, results in deposition of the dissolved
substance on the substrate. In fact, this deposition process can only
be controlled through controlling the dynamics of the carrier liquid film, and, in 
particular, the evolution of the morphology of the triple line. In
a recent study, DNA molecules were deposited in a highly ordered array by
dissolving them in a solvent and letting the solvent dewet
a micropillar-structured surface \cite{Lin}.

The dynamics of wetting on flat solid and liquid surfaces is quite well
understood \cite{Xue, Gau}, however, despite its applicability, only a few
experiments were performed on inhomogeneous, either patterned or
disordered surfaces \cite{cubaud1, cubaud2, cubaud3, fermigier4, clotet, duprat}, 
while the dynamics of a receeding contact line remains almost unexplored.
In spite of the apparent simplicity of the phenomenon, there are
no simple, easily manageable models for describing it. Although in the
lubrication approximation the Navier-Stokes (or, in the highly viscous regime
the Stokes) equation reduces to two dimensions \cite{revmodphys}, the
numerical modeling of layers with large planar extent is still
computationally time consuming and cumbersome due to the discontinuities on
the liquid-solid and liquid-gas interfaces. These discontinuities are
tackled within the framework of phase-field models \cite{phase_field}, but  it remains 
unclear however, how substrate inhomogeneities would be introduced in such
models. It is also also unsettled how the actual dynamics of the layer is influenced by the
chosen particular form of the phase interface.

The continuous emergence of newer and newer schemes in the topic suggests that
the demand for a convenient approach for modeling thin liquid layers' dynamics
is still unsatisfied \cite{phase_field, sprittles, PRL_savva, du, beltrame,
liao}. Based on the revolutionary paper of J. F. Joanny and P. G. de Gennes on 
the perturbed contact line shape \cite{deGennes3}, a series of depinning type 
models were constructed that aimed to describe interface dynamics in presence 
of disorder \cite{damien1, damien2, damien3}. These models are not restricted 
to dewetting phenomena, as they apply to fracture front propagation or even magnetic 
domain wall motion. In the framework of these models, small deformations 
of the interface and a linear restoring force acting on the contact line resulting 
from a perturbative approach are considered. They are thus inherently linear, 
and the only source of nonlinearity is the disorder of the landscape they propagate in. 
Although they have had a great success in the sampling of the depinning transition 
and determination of various critical exponents \cite{tanguy1, tanguy2}, they have 
the drawback that they neither allow for large deformations, nor for local backward 
movement of the line. Consequently, they are unable to account for the 
tearing up of the dewetting film, which, in fact,  is a common phenomenon.

Our purpose here is precisely to address the question of large deformations and the 
eventual tearing of the film with an efficient and easily manageable model for the contact 
line motion. Our method works best for viscous, flat and extended droplets with small wetting angle. 
It is shown that in this regime, in contrast to the perturbative treatment 
\cite{deGennes3}, the line is soft and ductile, meaning that a localized perturbation of the 
line induces only short range forces. Considering a viscous regime, the line's equation of 
evolution becomes an overdamped one. In the following sections we  will describe this method in detail, we 
will show how to handle substrate inhomogeneities, and an application is presented.

\section{Basic concepts}

Let the upper surface of the contracting fluid layer be described by $z=z(x,y,t)$. Our approach is restricted to the description of large, flat layers in the
highly viscous regime, the same assumption that is made when deriving the
lubricant equations \cite{revmodphys}, i.e. $|\nabla z| \ll 1$. One further assumption we make is
that the relative change in the height of the droplet is small, therefore
its height is almost constant in time, $\partial z/ \partial t \to 0$. Under these considerations, the layer's
free energy has two terms. The first component is the joint contribution of the
well-known liquid-solid and liquid-gas (air) surface tensions. If the layer is
flat, its upper and lower surface areas are approximately equal, $S$. Denoting
by $\gamma_{XY}$ the appropriate surface tension coefficients, the surface energy
writes as:
\begin{equation} \label{surface1}
 U_{surface} = \gamma_{SL} S + \gamma_{LG} S = \gamma S
\end{equation}
The second contribution to the total free energy of the layer is the line
energy which occurs due to the unbalanced forces acting on the layer boundaries  on the molecules 
from the liquid-substrate-air triple interface. 
This is a curve with finite
thickness, thus this energy is comparable to the surface energy and it is
proportional to the length of the triple interface, $l$
\begin{equation}\label{line1}
 U_{line} = \alpha l,
\end{equation}
where $\alpha$ is the line tension coefficient.  Neither the interpretation of
$\alpha$, nor its measurement is straightforward, in fact, there is still less
consensus regarding its magnitude: values ranging from $10^{-11} N$ to
$10^{-6} N$ were measured or computed in various experiments and simulations
\cite{popescu, binder, drelich1, drelich2, tadmor}. 
The major difficulty arises from the fact that dewetting is often accompanied 
by a precursor layer with a much smaller thickness than the rest of the layer. 
In our case, in term (\ref{line1}) a contribution resulting from the layer's side surface 
has to be also considered. This yields an extra surface energy that is also proportional with $l$, 
consequently, we believe that an effective $\alpha$
has to be used instead. Therefore in calculations larger values than the presented range should
be used. In the case of a real two dimensional flow (for instance,
flow in a Hele-Shaw cell \cite{clotet, fermigier4}), the line tension is well
defined and it is clearly a result from the finite side surface of the layer
between the plates. For complete wetting, i.e. zero wetting angle, $\alpha = \pi
\gamma_{LG} h/2$, where $h$ is the distance between the plates of the Hele-Shaw
cell \cite{fermigier4}. Alternatively, if a quantitative upscaling of the
elastic type of energy introduced in \cite{deGennes3} was possible (properly
removing the third dimension from the model), it could provide the correct
expression for the line tension for sufficiently flat droplets, bounded by one
solid surface only. Such an expression however is not available, hence it
remains an open question. 

The total free energy of the system is the sum of these two
contributions: $U = U_{surface} + U_{line}$. 

Our approach is based on the fact that both the surface and the line
energies are functionals of the shape of the triple interface, which is a
one-dimensional curve. When inertial effects do not play an important role 
(the highly viscous, low Reynolds number regime), the
total energy of the system is uniquely defined by the shape of the contact
line, it is therefore enough to track solely its dynamics. 

In order to illustrate this, we consider a simple example:  the dynamics of a circular hole. 
Due to the symmetry of the problem, an analytically study is possible. 
From energy terms (\ref{surface1}) and (\ref{line1}) the forces
acting on the edge of the hole can be derived, which, due to symmetry
considerations act in the radial direction
\begin{equation}\label{f_surface}
 F_{surface} = - \frac{\partial U_{surface}}{\partial R} = - \frac{\partial
}{\partial R} (- \gamma \pi R^2) =2 \pi \gamma R,
\end{equation}
where $R$ is the radius of the hole. Similarly, the force resulting from the
line tension:
\begin{equation} \label{f_line}
 F_{line} = - \frac{\partial U_{line}}{\partial R} = - \frac{\partial
}{\partial R} (\alpha 2 \pi R) = - 2 \pi \alpha
\end{equation}
Assuming an overdamped motion of the edge of the
hole (the triple interface), the following equation of motion yields for its
radius:
\begin{equation}
 (F_{line} + F_{surface}) m = \frac{dR}{dt}
\end{equation}
In the above expression, $m$ is the mobility of the three-phase line and is
inversely proportional to its length, i.e. the longer the line, the more
sluggish it is: $m = m_0 l_0/ (2 \pi R)$, where $m_0$ is the mobility of a line segment of length $l_0$.  The equation of motion for the contact line is thus:
\begin{equation} \label{motion1}
 \left(\gamma - \frac{\alpha}{R} \right) m_0\ l_0= \frac{dR}{dt}
\end{equation}
It can be seen that the equilibrium radius of the hole is $R_0 =
\alpha/\gamma$ which is an intrinsic length scale of the system. For large radii ($R/R_0 \gg 1 $) the line energy can be neglected and the
velocity of the contact line is constant:
\begin{equation} \label{veloc1}
 \frac{dR}{dt} = \gamma m_0\ l_0
\end{equation}

Note that when $R$ is large, $R(t) \propto
t$, which is in complete concordance with previous results, for instance
\cite{Xue}.
So far the mobility of the triple interface has been introduced as a
phenomenological parameter which, in turn, defines the time-scale of the
problem. Considering the case when no
slippage of the interface occurs (the flow of the interface is a Poiseuille
flow), in previous studies similar results to  eq. (\ref{veloc1}) have been
derived for the radial velocity of the triple interface for a drying patch nucleated into a liquid film \cite{Xue, PRL_Redon, redon,brochard}:
\begin{equation} \label{veloc2}
 \frac{dR}{dt} = \frac{\theta_e ^3}{12 \sqrt{2} \ln(\theta_e l/b) \mu} \gamma
\end{equation}
where $\theta_e$ is the equilibrium contact angle, $l$ is the rim width, $b$ is
the extrapolation length (the distance from the rim at which the velocity
extrapolates to zero) and $\mu$ is the viscosity.
Comparing eq. (\ref{veloc1}) to eq. (\ref{veloc2}) one can identify the
mobility given now in terms of independently measurable
quantities that are now properties of the contact line:

\begin{equation} \label{mobil}
 m_0 \ l_0= \frac{\theta_e ^3}{12 \sqrt{2} \ln(\theta_e l/b) \mu}
\end{equation}

In case of a curve-like interface with parametric equation $\vec{r}=\vec{r}(\theta)$ (where $\theta$
is some arbitrary parameter), the equation of motion writes as
\begin{equation}\label{eqm3}
 \dot{\vec{r}}(\theta) = m[\vec{r}(\theta)]\ \cdot \vec{F}[\vec{r}(\theta)],
\end{equation}
hence the mobility and the force in this case are both functionals of the shape
of the interface.

\section{The simulation method}

\begin{figure}[h]
\begin{center}
\includegraphics[width=6cm]{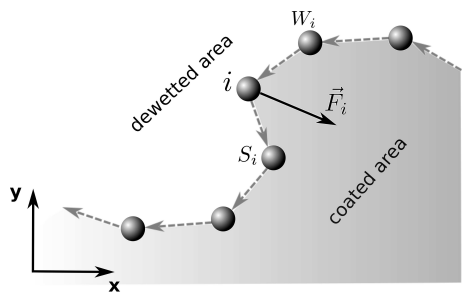}
\caption{\label{fig:model_sketch}Discretization of the contact line.}
\end{center}
\end{figure}

In order to model the dynamics of contact lines of arbitrary shape, numerical
methods are necessary. As a first step, the contact line is discretized
into {\em representative points}. After the contour is discretized, the
points are connected through directed line segments (vectors).
Each of the points "tracks" its previous and upcoming neighbors and, by
convention, the vectors are directed so that the liquid always lies on their
left hand side. Following the direction of the vectors connecting the points, a
directed chain is established. We denote by $S_i$ the index
of the ensuing point corresponding to point $i$ and by $W_i$ the point preceding $i$ (Fig.\ref{fig:model_sketch}).
In terms of the representative points' coordinates, the line and surface
tension energies write as:
\begin{eqnarray}
U_{line} & = & \alpha \sum_i \sqrt{(x_i-x_{S_i})^2 + (y_i-y_{S_i})^2} \\
U_{surface} & = &\gamma \frac{1}{2} \sum_i x_i y_{S_i} - x_{S_i}y_i
\end{eqnarray}
Once the energies are obtained, the forces acting on the representative points
are computed as $\vec{F}_i =- \nabla_i U$. In our two-dimensional approximation, the two components of this force are
\begin{eqnarray} \label{forces}
F_{ix}& = & -\frac{\partial U}{\partial x_i} = \\
     & = & - \alpha \left[ \frac{x_i-x_{W_i}}{d_{i, W_i}} +
\frac{x_i-x_{S_i}}{d_{i, S_i}}
\right] + \gamma(y_{S_i} - y_{W_i})  \nonumber \\
F_{iy}& = & -\frac{\partial U}{\partial y_i} = \\
     & = & - \alpha \left[ \frac{y_i-y_{W_i}}{d_{i, W_i}} +
\frac{y_i-y_{S_i}}{d_{i, S_i}}
\right] - \gamma(x_{S_i} - x_{W_i}) \nonumber,
\end{eqnarray}
where $d_{k,l}$ is the distance between points $k$ and $l$.
It can be readily seen that each point interacts with its nearest neighbors
only, therefore a molecular dynamics type simulation is suitable for
investigating their dynamics. We emphasize that the localized nature of the forces is a direct consequence of our primary hypothesis, i.e. the droplet is flat and its height profile does not change significantly during the movement of the contact line. Either at lower scales, where the fine structure of the contact line becomes relevant or in the case of non-flat droplets the Green function of the contact line (its response to a localized perturbation) is of long-range nature. As mentioned in the introduction, a perturbative treatment for small deformations of the contact line is described in Ref. \cite{deGennes3}, while the propagation of such lines in random media resulting in depinning transition and a consequent advancing accompanied by avalanches are extensively studied in Ref. \cite{damien2} and \cite{damien3}.
For the present case, we stick to the lubricant approximation, thus proceed with eq. (\ref{forces}). 
The overdamped equation of motion for
the points is:
\begin{eqnarray} \label{eqm1}
\dot{\vec{r}_i}  =  m_i \vec{F}_{i} 
\end{eqnarray}
The mobility $m_i$ associated to point $i$ is inversely proportional to the
length element of the respective point on the triple interface:
\begin{equation}
 m_i = m_0 \frac{2 d_{max}}{d_{i, S_i} + d_{i, W_i}},
\end{equation}
where we remind  that $d_{i, S_i}$ is the distance between point $i$ and its upcoming neighbor, while $d_{i, W_i}$ is the distance between point $i$ and its previous neighbor. During their dynamics, the representative points will approach or move away from each other. In order to preserve numerical accuracy, their density on the triple line should be kept constant. Imposing a constant density however, is incompatible with the movement of the individual points, therefore, an optimal fluctuation around an average value is necessary. This issue is solved by inserting a new point between two neighboring points whenever they move farther than a predefined distance $d_{max}$. In case they come closer than another predefined distance $d_{min}$, one of the points is removed. As a rule of thumb, we 
consider $d_{min} = 0.8\ d_{max}/2$, which ensures that no insertion is necessary right after a removal. With this choice, $m_0$ is then the mobility of one line segment. Note that continuous indexing of neighboring points is not possible due 
to the repeated insertions and removals.

\begin{figure*}[h]
\begin{center}
\includegraphics[width=14cm]{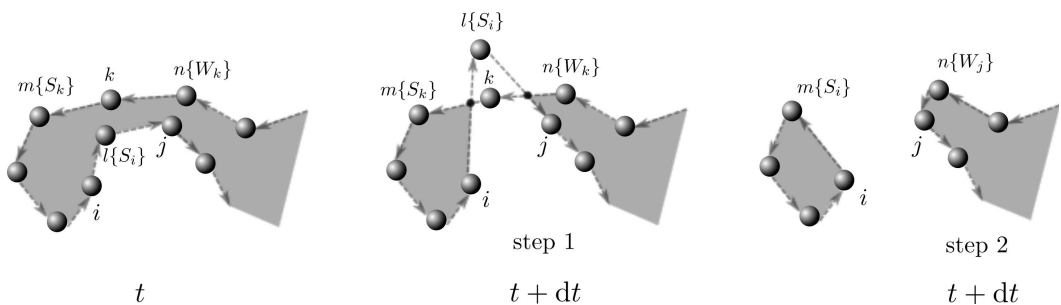}
\caption{\label{fig:tear_sketch} The reconnection mechanism for the tearing of the layer. The label values in parenthesis indicates the succeeding order in the oriented chain. Please note that  step 1 and step 2 are made in the same time moment. }
\end{center}
\end{figure*}

Whenever two segments intersect, the points are reconnected such that the line breaks up, hence allowing for tearing the layer.
The used reconnection mechanism is sketched on Figure \ref{fig:tear_sketch}. 

\section{Inhomogeneities}
Similarly to previous descriptions, one may introduce inhomogeneities of the
substrate in terms of pinning points. Whenever the contact line hits a pinning
point, it is blocked as long as the force acting on it does not reach a given
threshold. Eq. \ref{eqm3} then modifies to:
\begin{eqnarray} \label{eqm2}
\dot{\vec{r}}(\theta)  =  m[\vec{r}(\theta)] \cdot \left(\vec{F}[\vec{r}(\theta)] +\vec{F}_{pin}[\vec{r}(\theta)] \right)
\end{eqnarray}
where $\vec{F}_{pin}$ is the pinning force resulting from 
inhomogeneities:
\begin{equation}
        \vec{F}_{pin}(\vec{r}) = \begin{cases}
                        - \eta(\vec{r}) \frac{\vec{F}(\vec{r})}{\mid \vec{F}(\vec{r}) \mid} & \text{ if } \mid \vec{F}(\vec{r}) \mid > \eta({\vec{r}}) \\
                        -\vec{F}(\vec{r}) & \text{ if } \mid \vec{F}(\vec{r}) \mid \le \eta({\vec{r}})
                    \end{cases}
\end{equation}
Here $\eta(\vec{r})$ will characterize the pinning strength at site with
position at $\vec{r}$. In case of point-like inhomogeneities, localized at spatial
coordinates $\vec{r}_k$
\begin{equation}
        \eta(\vec{r}) = \begin{cases}
                        \eta_k & \text{ if } \vec{r}= \vec{r}_k \\
                        0 &  \vec{r}\ne \vec{r}_k,
                    \end{cases}
\end{equation}

where $\eta_k$ are the thresholds of the pinning points. In the 
followings, spatially uniformly distributed and
uncorrelated inhomogeneities are considered. 
For simplicity reasons, the $\eta_k$ threshold values are considered also uniformly and
uncorrelatedly distributed on the  $[0, \eta_0)$ interval.

The concentration of the point-like inhomogeneities is
\begin{equation}
c = \lim_{S \to \infty} \frac{1}{S} \int_S \sum_k \delta(\vec{r}-\vec{r}_k) d
\vec{r},
\end{equation}
while their average distance is given by $L_0=1/\sqrt{c}$.

Note that the disorder is quenched, which means that in principle their
positions would have to be generated and fixed right from the
beginning of the simulation. The line segments have to be tested at any instant of the simulation, 
whether they cross any of the pinning points, a procedure which is extremely time consuming. In
order to avoid this, a simplified procedure is used to generate pinning points on
the run, yet preserving their statistical properties.
\begin{figure}[h]
\begin{center}
\includegraphics[width=6cm]{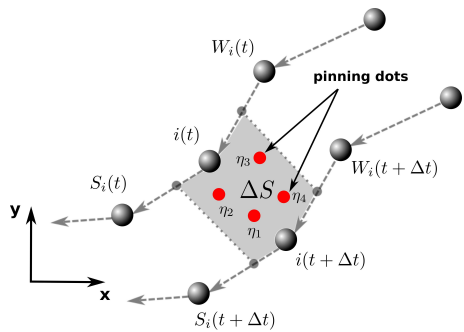}
\caption{\label{fig:pinning}(color online) Handling the substrate inhomogeneities. In this example, the line segment corresponding to point $i$ crosses $4$ pinning centers, each with its own threshold. The effective threshold experienced by point $i$ is the largest one out of those $4$. The pinning points are considered point-like, with no planar extension.}
\end{center}
\end{figure}

If the line segment belonging to point $i$ sweeps a small area $\Delta S$
within a time interval $\Delta t$ (Fig. \ref{fig:pinning}), the probability of
finding exactly $n$ pinning points within that area has a Poisson distribution:
\begin{equation}\label{poisson}
 P(n) = \frac{1}{n!} (c \Delta S)^n \exp(-c \Delta S)
\end{equation}
Since the pinning is related to thresholds, whenever the line segment crosses $n$ pinning points, with thresholds $\{\eta_1, \eta_2, ..., \eta_n\}$, it will experience an effective threshold which is the maximum of all the thresholds of the points within $\Delta S$:
\begin{equation}
 \eta_{eff} = \max \{\eta_1, \eta_2, ..., \eta_n\}
\end{equation}
Bearing in mind that $\eta_k$ is uniformly distributed on the  $[0, \eta_0)$ interval, the probability
distribution of the maximum is given by:
\begin{equation} \label{maxd}
 P(\eta_{eff}|n) =
 n\ \eta_{eff}^{n-1}\ \eta_0^{n-2} \ \  \mbox{where}\ \eta_{eff}<\eta_0 \\
\end{equation}

At every time step, for each site, the number of pinning points is drawn according to the (\ref{poisson}) distribution, while the thresholds is generated according to the  (\ref{maxd}) distribution.

\section{Application: a soft dewetting line under quenched disorder}

As application to the previously discussed method, we will study the dynamics and topology of a moving dewetting line on a substrate with uniformly distributed quenched disorders.  
Disorders act as pinning centers,  and we consider them point-like with the statistical properties  described in the previous section. 
The initial state of the interface is a straight line along the 
$x$ axis ($y(t=0)=0$), and the liquid is considered to be under this line in the $y<0$ semiplane. 
Periodic boundary conditions are imposed along the $x$ axis, hence while the liquid contracts,
the contact line moves towards the negative $y$ direction. After a transient period, the line reaches a dynamic equilibrium state, in which its statistical properties are stationary. 

$R_0 =\alpha/\gamma$ is chosen as the unit length of the simulation.
 All the lengths are then expressed in terms of dimensionless coordinates $\widetilde{\vec{r}}={\vec{r}}/R_0$.
Let us introduce $R_1=\eta_0/\gamma$, which would correspond to a flat line element subjected to a capillary force that 
would move it over a pinning dot with threshold $\eta_0$.  Its dimensionless form is $\widetilde{R}_1=R_1/R_0$. 
The dimensionless time is $\widetilde{t} = \gamma m_0 t$. The equation of motion (\ref{eqm2}) can then be rewritten 
in terms of these dimensionless quantities which leaves us with two parameters only: the length scale $\widetilde{R}_1$ defined 
by the amplitude of the inhomogeneity thresholds and the length scale $\widetilde{L}_0 = L_0/R_0$ defined by their concentration. 
Consequently, the dynamics of the line is a result of the competition between
these two length scales.

Simulations were carried out for a system length along the $x$ direction
$\widetilde{L}_x = 160$, representative points distance $\widetilde{d}_{max} = 0.2$ 
and a time step $\Delta \widetilde{t} = 10^{-3}$. As mentioned previously,
tearing up of the layer is possible, however, the resulting droplets (droplets
left behind as a result of tearing) are disregarded as they have no more
influence on the dynamics of the main line. Wherever it was meaningful, an
ensemble average over $10$ ensembles (independent runs) was considered.

\section{Results and discussion}

\begin{figure*}[h]
\begin{center}
\includegraphics[width=14cm]{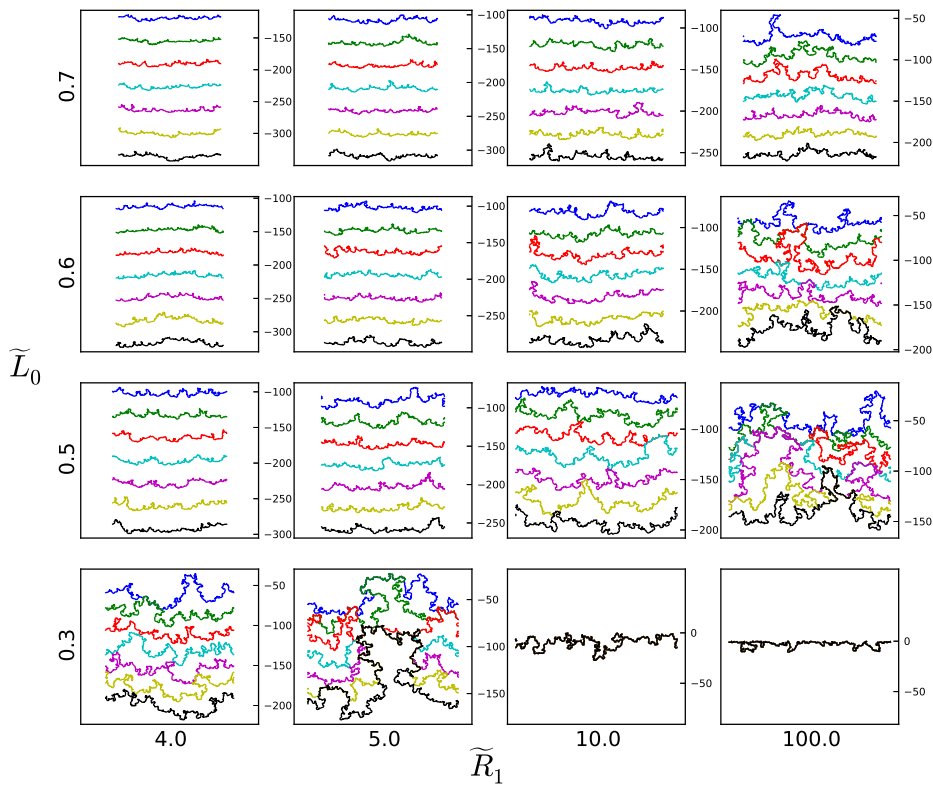}
\caption{\label{fig:coordinates} (color online) Contact line morphology for equally spaced time moments (plots in the $x-y$ plane). Evolution of the interface is from top to bottom (from the blue line to the black one). The inset graphs from left to right correspond to increasing $\widetilde{R}_1$ values (indicated in the horizontal direction), while from bottom to top we consider increasing $\widetilde{L}_0$ values (indicated in the vertical direction). The roughness and velocity fluctuations increase, long range correlation and large deformation develops as the system approaches the depinning transition. In the two bottom-right cases, after sweeping a finite distance, the line is pinned. In order to better visualize every position of the line within the desired interval, different scales on the $y$ axis have been used. The scale in the $x$ direction is always $200$ units}
\end{center}
\end{figure*}

We present now the results obtained for the dynamics of the model system described in the previous section (section V.).
First, we study qualitatively the dynamics of the interface.
Fig. \ref{fig:coordinates} shows the time evolution of the contact line for various parameters $\widetilde{R}_1$ and $\widetilde{L}_0 $. 
As the line's average velocity decreases, i. e. as it approaches the depinning transition, its length and roughness increases. One will observe that the contact line reaches a statistically stable conformation, and
its shapes are in good qualitative agreement with the experiments carried out by Clotet et al.\cite{clotet} and Paterson et al. \cite{fermigier4, paterson} in a Hele-Shaw cell, although both experiments were carried out for wetting on disordered substrates, i.e. the opposite dynamics of the contact line.

\begin{figure}[h]
\begin{center}
\includegraphics[width=8cm]{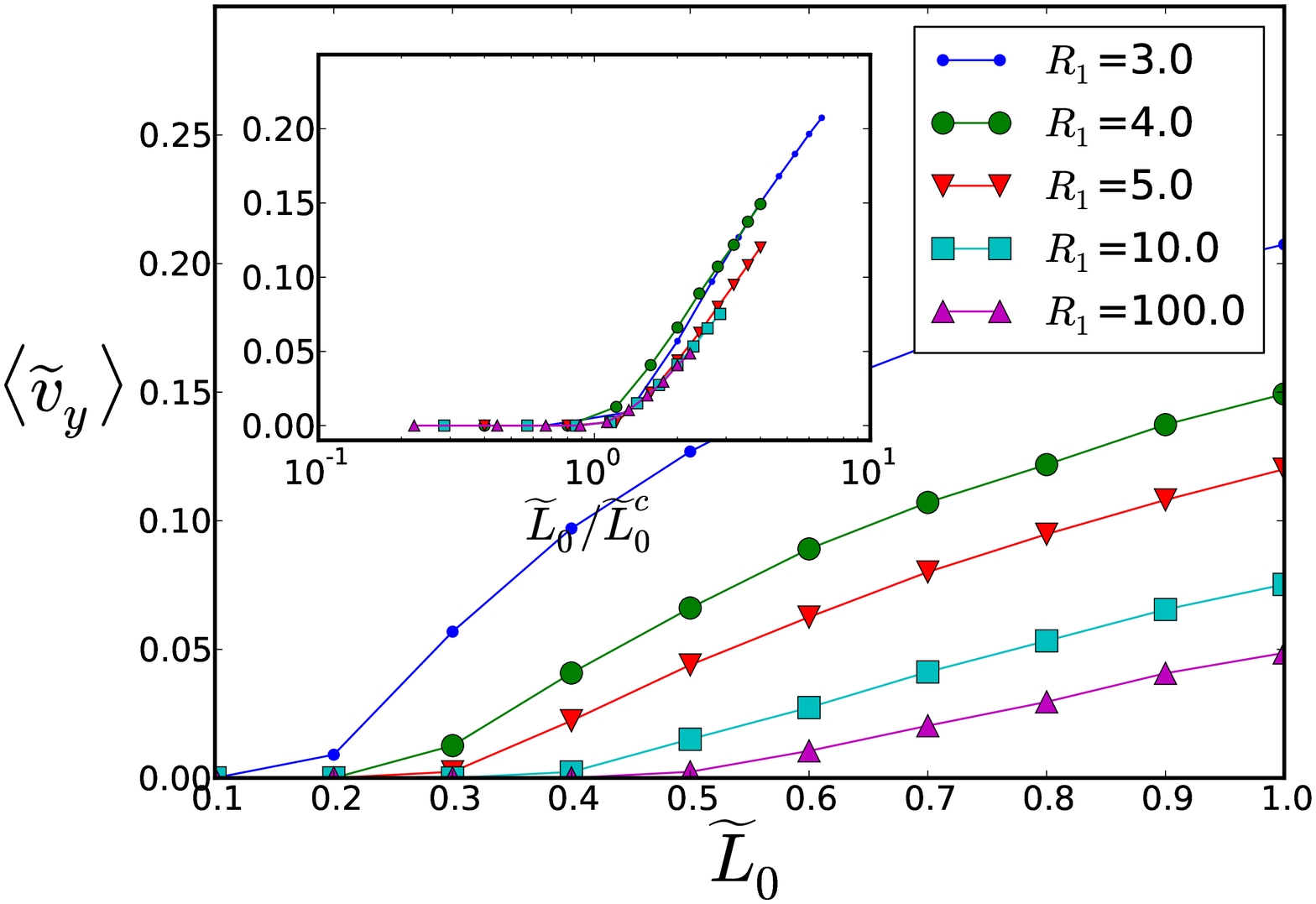}
\caption{\label{fig:velocity_vs_L0} (color online)  Mean velocity of the interface along the $y$ axis, in the stationary regime, as a function of $\widetilde{L}_0$ for different $\widetilde{R}_1$ parameters. The inset shows the mean velocity as a function of $\widetilde{L}_0 /\widetilde{L}^c_0$. A reasonable collapse is obtained.}
\end{center}
\end{figure}

In the dynamic equilibrium (stationary regime of the moving interface), 
the mean velocity of the interface along the $y$ direction presents a nontrivial, phase-transition like
behavior as a function of $\widetilde{L}_0$. There is a critical concentration, below which the line is 
depinned (Fig. \ref{fig:velocity_vs_L0}) and this is what we call {\em depinning transition}. 
 
 This critical concentration (or, the associated length $\widetilde{L}^c_0$) depends on the pinning strength.  
 From Fig. \ref{fig:velocity_vs_L0} we also learn that 
$\widetilde{L}^c_0$ increases with $\widetilde{R}_1$ and converges to
$\widetilde{L}^c_0 = 1/2 \pm 0.1$ as $\widetilde{R}_1 \to \infty$. This value is
significantly lower than $\widetilde{L}^c_0 = 2$, which would be the critical
length for a regular array of defects with infinite strength that would prevent tearing.
Collective trapping of parts of the
contact line thus is possible if the distance between the neighboring defects is
less than $2$. The existence of such a  threshold, lower than $\widetilde{L}^c_0 = 2$ has been shown
experimentally \cite{paterson}, however, since the experiment was carried out in
gravity, its value is related to the capillary length. In our case, the obtained lower
limit is merely a consequence of the competition between the line and surface
tensions and the value $\widetilde{L}^c_0 = 1/2$ is thus a consequence of the
underlying disorder. It is related to the percolation of the contact line
between the localized defects. As it is expected for a critical behavior, the mean velocity curves have a reasonable collapse if they are plotted as a function of $\widetilde{L}_0 /\widetilde{L}^c_0$.
The inset in Figure \ref{fig:velocity_vs_L0} shows the results in such sense.

\begin{figure}[h]
\begin{center}
\includegraphics[width=8cm]{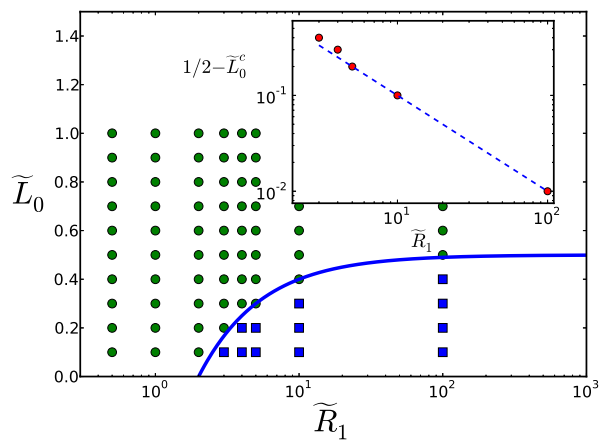}
\caption{\label{fig:phase_diagram} (color online) Phase diagram of the contact line in the $(\widetilde{R}_1 , \widetilde{L}_0)$ parameter space. Symbols indicate parameter values at which simulations were performed. Blue squares indicate the obtained pinning phase, green dots the depinning phase. The inset derived from the separation points shows that the two phases 
are delimited by the curve  
$\widetilde{L}^c_0 = 1/2 - \widetilde{R}_1^{-1}$, the dashed line indicating a slope $-1$. Please note the logarithmic scales for the inset graph.}
\end{center}
\end{figure}

Although the number of the simulated data points was rather limited for this purpose, we made an attempt to find the 
$\widetilde{L}^c_0 =  \widetilde{L}^c_0(\widetilde{R}_1)$ dependence. We considered the mesh illustrated on Fig. \ref{fig:phase_diagram} in the $\widetilde{R}_1-\widetilde{L}_0$  plane to detect the 
occurrence of the depinning transition.  The inset in Fig. \ref{fig:phase_diagram} shows that 
$\widetilde{L}^c_0 = 1/2 - \widetilde{R}_1^{-1}$ is a reasonable fit for describing the boundary between the two phases in the mapped region. Interestingly, this fit suggests that for $\widetilde{R}_1 < 2$ a total pinning is not possible.

For the high inhomogeneity and low threshold regime ($\widetilde{L}_0 \ll 1$, $\widetilde{R}_1 \ll 1$), one would expect the possibility of a classical depinning transition, with small deformations of the contact line. Interestingly however, we could not observe such a transition, even for extremely low values of $\widetilde{L}_0$ and $\widetilde{R}_1$. In their experiments, Duprat et al. \cite{duprat} investigated the depinning of a wetting contact line from an individual defect. They reported that depending on the pinning strength the contact line either jumped off the defect or completely wetted it, and advanced by tearing up and leaving an air hole behind. For individual or localized group of inhomogeneities we observed the same behavior, however, it turned out to be impossible to recover a collective depinning transition without the tearing up of the film.
This is probably the result of the high ductility of the contact line. The classical depinning transition  occurs due to the competition between disorder and long range elastic restoring forces \cite{kardar}, while in our case, we lack the long range part, therefore, we encounter a new transition, which is mainly governed by large deformations and tearing up of the layer. 
In the experiments of Paterson and Fermigier \cite{paterson}, the authors distinguish between strong and weak pinning as a function of the spatial distribution of the inhomogeneities. In the strong pinning case, defects were spread randomly and uniformly over the whole surface, while in the weak pinning case, they were spread by positioning randomly only one defect in each unit cell of a larger square lattice, hence obtaining a more homogeneous pattern. 
For the same defect concentration, the second case results in smaller average distance $\widetilde{L}_0$ between the defects. The observation that in the strong pinning case (small $\widetilde{L}_0$) the contact line breaks up, and in the weak pinning case (large $\widetilde{L}_0$) it advances with a rather smooth shape, is compatible with our simulation results, even though we tuned $\widetilde{L}_0$ by changing the defect concentration rather than changing their distribution or correlation.

\begin{figure}[h]
\begin{center}
\includegraphics[width=8cm]{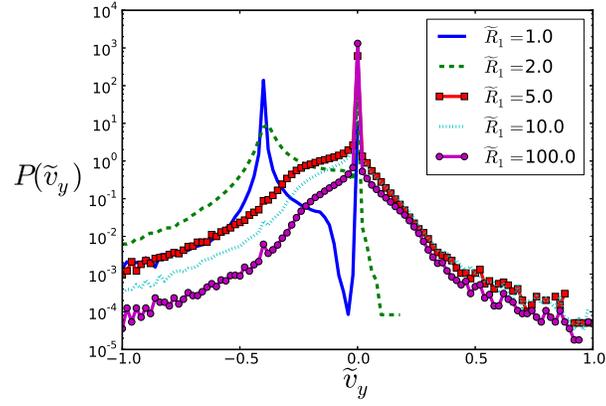}
\caption{\label{fig:velocities} (color online) Distribution of the $y$ component of velocities along the contact line for $\widetilde{L}_0=0.5$. Note that when we approach the depinning transition ($\widetilde{R}_1\to 10^2$) a considerable local backward movement ($v_y>0$) of the interface occurs. Also, far from the transition ($\widetilde{R}_1 \ll 10^2$), the pinned part of the line is quite well separated from the moving part.}
\end{center}
\end{figure}

Another major difference compared to classical depinning models is that in our system local backward movements of the interface may appear, and, indeed, approaching the transition, positive velocities of the representative points occur, which plays an important role in the roughening mechanism. Figure \ref{fig:velocities} shows how the distribution of the velocity components in the $y$ direction 
changes as we approach the transition point.  
Far from the transition point we experience an almost bimodal distribution (one peak corresponding to the unpinned part, while the other one, at zero, to the pinned part),  while close to it
we obtain an almost zero-averaged symmetric distribution. Clearly, it is due to the slight asymmetry that the contact line moves forward on average. 

\begin{figure}[h]
\begin{center}
\includegraphics[width=8cm]{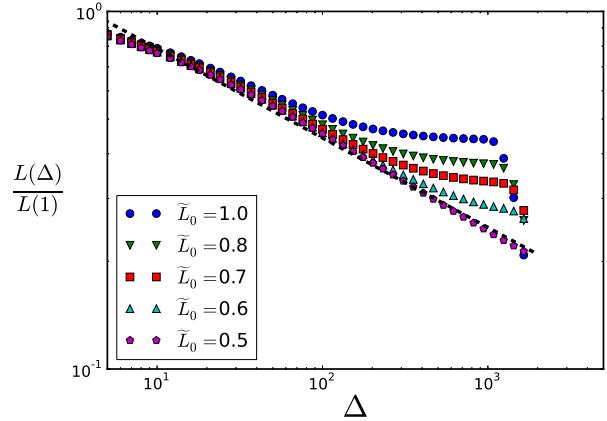}
\caption{\label{fig:L_delta} (color online) Development of the scale-free morphology as the system approaches the critical state. The normalized length $L(\Delta)$ of the contact line as a function of $\Delta$ (see the text for the definitions).  
Results for $\widetilde{R}_1=10^2$ and different values of $\widetilde{L}_0$. The dashed line is a guide for the eye, and has a slope $-0.25$. A natural upper cutoff arises due to the finite system size, and a lower cutoff from the discretization.}
\end{center}
\end{figure}

In order to quantify the morphology of the contact line around the transition, we performed a classical rasterization analysis. The length of the contact line $L$ was measured by taking into account only every $\Delta^{th}$ representative point, and the scaling of $L$ with respect to $\Delta$ was investigated. This means that for $\Delta=1$, $L$ is computed by adding up the distance between each nearest neighboring point, for $\Delta=2$ by summing the distance between each second neighbor points and so on, hence the length of the curve is approximated at different precisions. Figure \ref{fig:L_delta} shows that as the system approaches the depinning transition, the scaling converges to a power law, $L(\Delta) \propto \Delta^{-1/4}$. This suggests a fractal-like structure and a scale-free morphology with a diverging total length as $\Delta$ decreases. This is again a direct consequence of the undergoing phase transition.

\begin{figure}[h]
\begin{center}
\includegraphics[width=8cm]{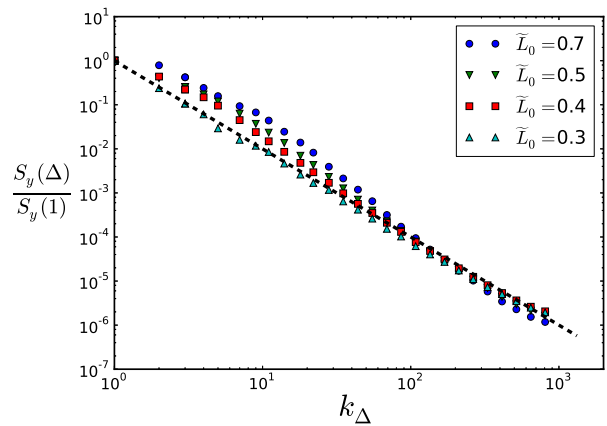}
\caption{\label{fig:S_delta} (color online) Development of the scale-free morphology as the system approaches the critical state. 
The structure factor $S_y(k_\Delta)$ as a function of $\Delta$ (see the text for definitions). 
Results for $\widetilde{R}_1=5.0$ and different values of $\widetilde{L}_0$. The dashed line has a slope $-2.0$ and the range $ 1 \le \Delta<2048$  was used for the Fourier transform.}
\end{center}
\end{figure}

Since $\Delta$ can be used to parametrize the contact line $(x(\Delta), y(\Delta))$, further information concerning its shape can be extracted by investigating the structure factor $S_y(k_\Delta)$ defined as the power spectrum of $y(\Delta)$:  $S_y(k_\Delta) =| \hat{y}(\Delta)|^2$ where $\hat{y}(\Delta)$ is the Fourier transform of $y(\Delta)$. Figure \ref{fig:S_delta} shows the convergence of $S_y(k_\Delta)$ to a power law in the vicinity of the transition point: $S_y(k_\Delta) \propto k_\Delta^{-2}$. This suggests again the scale-free, fractal-like shape for the interface. As expected, the main difference between the various curves $S_y(k_\Delta)$ arises from the low frequency, hence large wavelength values, showing that long range correlation develops close to the transition point.

\begin{figure}[h]
\begin{center}
\includegraphics[width=8cm]{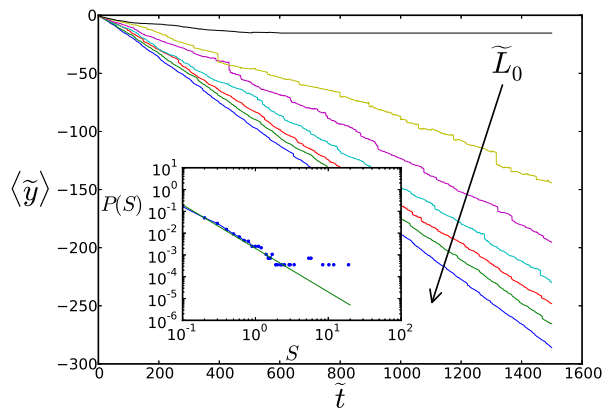}
\caption{\label{fig:y_avg} (color online) Average position of the contact line as a function of time, for 
$\widetilde{R}_1=10^2$ and $\widetilde{L}_0 = \{0.4, 0.5, 0.6, 0.7, 0.8, 0.9, 1.0 \}$. The arrow indicates increasing values of $\widetilde{L}_0$.  Note how fluctuations increase as approaching the transition point 
and the dynamics becomes intermittent. The inset shows the avalanche size distribution for $\widetilde{R}_1=10^2$ and  $\widetilde{L}_0 = 0.5$, while the solid line has a slope: $-2.0$.}
\end{center}
\end{figure}
The average position of the contact line was also followed as a function of time. Results for a  fixed $\widetilde{R}_1=10^2$  value and a wide range of  $\widetilde{L}_0$ values are plotted on Figure \ref{fig:y_avg}.  When approaching the critical point, fluctuations increase and the sudden jumps in the average position become more and more dominating. These jumps are the result of either the slip of the contact line over individual defects or the tearing up of the layer. Analogously to jumps in the magnetization (Barkhausen noise), these jumps are termed avalanches, since the average position of the line is governed by fast slips. Close to the transition, the sizes of the jumps exhibit a power-law distribution with an exponent $-2$ (inset of Fig. \ref{fig:y_avg}). Our results along this line are however modest (the scaling is on an interval less than two orders of magnitude), due to the lack of statistics for the large avalanche sizes. It is important to note however that experimental data presented in Ref 
\cite{moulinet} clearly shows values around -2, giving thus some confidence to the results of our model. 

\section{Conclusions}
A novel and efficient, off-lattice molecular dynamics type simulation has been introduced in order to
investigate the dynamics of thin and viscous liquid layers, dewetting on inhomogeneous surfaces. 
By using this simulation method the existence of an unusual depinning transition was 
revealed. This transition is governed by large deformations of the interface and the 
breaking up of the layer.  The two-dimensional parameter space of the investigated system was thoroughly explored, and the obtained results
were discussed in view of  available experimental observations. 
We learned that the contact line's dynamics is a result of an interplay between the capillary forces and 
the substrate disorder, however, with the appropriately introduced adimensional form, both relevant 
parameters are related to the inhomogeneities. In such an approach, the universal properties of the contact 
line can be viewed as a result of the competition between the inhomogeneities strength and their density.
The difference between the dynamics of a receeding and an advancing contact line (dewetting vs. wetting), other than the contact angle hysteresis, remains an open question and could be investigated in the future by introducing pressure in our model.

\section{Acknowledgments}
The work of Z. Neda was supported by the Romanian IDEAS  PN-II-ID-PCE-2012-4-0470 research grant.
The research of B. Tyukodi was supported by the European Union and the State of Hungary,
co-financed by the European Social Fund in the framework of T\'AMOP 4.2.4.A/2-11-1-2012-0001 
‘National Excellence Program’. We are grateful for Damien Vandembroucq and Etienne Barthel for the careful reading and relevant comments on the manuscript.

\end{document}